\documentclass[twocolumn]{svjour3} 

\usepackage[utf8]{inputenc}
\usepackage{graphicx}
\usepackage{float}
\usepackage{psfrag}
\usepackage{pstricks}
\usepackage{tikz}
\usepackage{upgreek}
\usepackage{subcaption}
\usepackage{amsbsy}
\usepackage{amssymb}
\usepackage{pdfpages}
\usepackage{amsmath}
\usepackage{wasysym}
\usepackage[round]{natbib}
\usepackage{picture}
\usepackage[]{hyperref}
\usepackage{etoolbox}
\usepackage{color}
\usepackage{xcolor}
\definecolor{mypurp}{RGB}{148,0,211}
\usepackage{soul}
\hypersetup{colorlinks,linkcolor={blue},citecolor={blue},urlcolor={red}} 
\graphicspath{ {Figures/} }  

\title{Characterizing the surface texture of a dense suspension undergoing dynamic jamming}

\author{Olav Rømcke     \and
       Ivo R. Peters    \and
       R. Jason Hearst
}

\institute{ Olav Rømcke,  R. Jason Hearst\at
            Department of Energy and Process Engineering\\
            Norwegian University of Science and Technology\\
            Kolbjørn Hejes vei 2, NO-7491 Trondheim, Norway\\
            \email{jason.hearst@ntnu.no}  
            \and
            Ivo R. Peters \at
            Faculty of Engineering and Physical Sciences\\
            University of Southampton \\
            Highfield, Southampton SO17 1BJ, UK
            \and
}

\date{Received: date / Accepted: date}

\begin{document}


\maketitle
\section*{Abstract}
Measurements of the surface velocity and surface texture of a freely propagating shear jamming front in a dense suspension are compared. The velocity fields are captured with particle image velocimetry (PIV), while the surface texture is captured in a separated experiment by observing a direct reflection on the suspension surface with high-speed cameras. A method for quantifying the surface features and their orientation is presented based on the fast Fourier transform of localised windows. The region that exhibits strong surface features corresponds to the the solid-like jammed region identified via the PIV measurements. Moreover, the surface features within the jammed region are predominantly oriented in the same direction as the eigenvectors of the strain tensor. Thus, from images of the free surface, our analysis is able to show that the surface texture contains information on the principle strain directions and the propagation of the jamming front. 

\keywords{}


\section{Introduction}
\label{sec:Introduction}

Suspensions of hard spheres in a Newtonian fluid are known to jam at a critical volume fraction \citep{Krieger1972}. That is, beyond a critical concentration of particles, flow ceases and a finite yield stress is observed \cite{Brown2014}. However, for some suspensions, such as cornstarch and water, a jammed state is accessible for volume fractions lower than the critical volume fraction when stress is applied \citep{Wyart2014, Peters2016}. This rather counter intuitive phenomenon is called dynamic jamming or shear jamming, where the suspension appears fluid-like at low stress, but shear thickens and even jams with sufficiently high stress. As such, a sudden impact causes the suspension to jam \citep{Waitukaitis2012, Jerome2016}, which explains how it is possible to stay afloat while running over a cornstarch suspension \citep{Mukhopadhyay2018, Baumgarten2019}.

The assumption of smooth, force free particles in non-Brownian, non-inertial systems, where viscosity is only a function of the volume fraction, $\phi$ \citep{Stickel2005}, does not capture this behaviour. In real suspensions particle-particle interactions are an important contributor to the observed behaviour \citep{Lin2015, Gadala1980, Brown2014}. Between particles, it has been identified that friction \citep{Mari2014, Singh2018, Sivadasan2019, Tapia2019, Madraki2017, Fernandez2013} and repulsive forces \citep{Brown2014, James2018, Guy2015} are underlying mechanisms for understanding this phenomenon. With a sufficient amount of applied stress, the particles overcome the repulsive force, and are brought into frictional contact. At sufficiently high particle concentrations, the contacts form a network capable of supporting the applied stresses.

In sufficiently large domains, the transition from fluid-like to solid-like is observed as a front of high shear rate that propagates from the perturbing body through the suspension and leaves a jammed state in its wake \citep{Waitukaitis2012, Peters2014, Han2016, Peters2016, Majumdar2017, Han2018, Han2019, Baumgarten2019, Romcke2021a}. In these works, the jamming front is defined by the velocity contour at half the velocity of the perturbing body, i.e., $0.5U_c$. A normalized front propagation factor is used to quantify how fast the front moves, defined by the relation between the speed of the $0.5U_c$-contour and the perturbing body. The front propagation factor is observed to increase with volume fraction and is independent of perturbing speed for sufficiently high velocities \citep{Han2016, Romcke2021a}. This phenomenology is caused by an intrinsic strain \citep{Han2019b, Baumgarten2019} which is needed in order for the material to build a frictional contact network capable of supporting the applied stresses. The strain level decreases with increasing volume fraction and has an inverse relationship with the front propagation factor \citep{Han2019b}.

Most measurement set-ups employed to investigate this problem have a free surface, and as such, the effects of the free surface have been identified as an important question in suspension flow \citep{Denn2018}. One interesting feature is the existence of two statically stable states \citep{Cates2005, Cates2014} known as granulation. The material can exist as a flowable droplet with a shiny surface, or in a stressed state as a jammed, pasty granule upheld by capillary forces. A closely linked observable surface feature in dense suspension flow is dilation \citep{Brown2012, Jerome2016, Maharjan2021}. For a sufficiently dense suspension, the granular structure expands under shear, which sets up a suction in the liquid phase. Dilation can thus be observed at the free surface as a transition from reflective to matte as individual particles protrude through the liquid-air interface. Dilation is associated with a large increase in stress \citep{Maharjan2021}, and coupled with the suspending fluid pressure \citep{Jerome2016} is able to explain the fluid-solid transition observed in impact experiments with a solid sphere. For a shear jamming front under extension, a reflective-matte transition is observed when the front interacts with the wall \citep{Majumdar2017}. 

A corrugated free surface has been reported for a wide range of particle sizes and packing fractions and in several experimental setups \citep{Loimer2002, Timberlake2005, Singh2006, Kumar2016}. In the inclined plane experiment by \citet{Timberlake2005}, two dimensional (2D) power spectra of free surface images indicate that the features exhibit anisotropy, specifically, the corrugations are shorter in the flow direction. Probably more applicable to the work herein is that of \citet{Loimer2002} who conducted experiments in an approximately 2D belt driven shear cell with the free surface normal in the vorticity direction. Power spectra in the flow and gradient directions, respectively, also indicate anisotropy. However, how these features appear in the full 2D power spectra remains unclear. The deformation of the free surface is a result of shear induced normal stresses \citep{Timberlake2005, Brown2012}, typically observed in dense suspensions \citep{Brown2014, Guazzelli2018, Denn2018}. That is, upon shearing, the material responds with a force normal to the confining boundary. Although several experiments investigating the dynamic jamming front phenomenon exhibit a large free surface \citep{Peters2014, Han2018, Romcke2021a}, few studies have dedicated attention to the developing surface texture as the front propagates through the suspension \citep{Allen2018}. 


In this work, we present observations of the free surface texture as the jamming front propagates unimpeded through the suspension. The aim of the method presented here is to draw quantifiable information from high-speed photographs of the free surface alone, without the need for more complex techniques, e.g., particle image velocimetry (PIV). The result from the free surface images is compared with the velocity field, front propagation and the strain tensor acquired from PIV measurements.


\section{Experimental procedure}
\label{sec:DataExp}


The data used here are collected from two different experiments. A single cylinder is traversed through a layer of cornstarch and sucrose-water suspension. First, as a reference, the free surface was seeded with black pepper. High-speed images of the suspension surface were captured under indirect lighting. PIV was conducted on these particle images, resulting in a time resolved velocity field. Secondly, by minor adjustments to the set-up, we record the free surface. In this case, the suspension is not seeded, while the camera was positioned such that it observed a direct reflection on the free surface, enhancing the visibility of any surface features. 

\begin{figure}
  \centering
  \subcaptionbox{\label{fig:SETUP:PIV}}{\includegraphics[width=0.67\columnwidth]{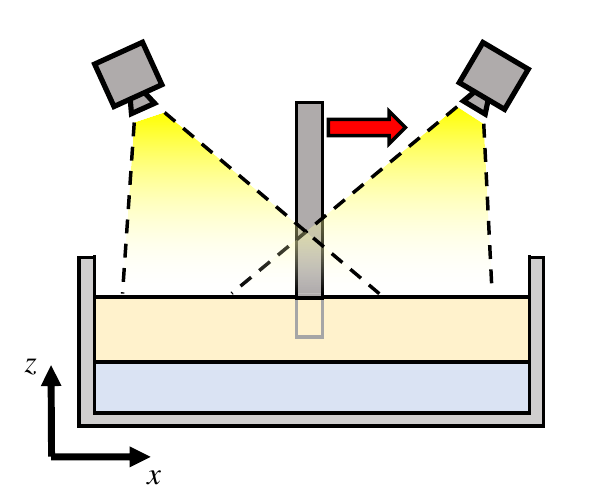}}\\\vspace{0.5em}
  \subcaptionbox{\label{fig:SETUP:TEX}}{\includegraphics[width=0.67\columnwidth]{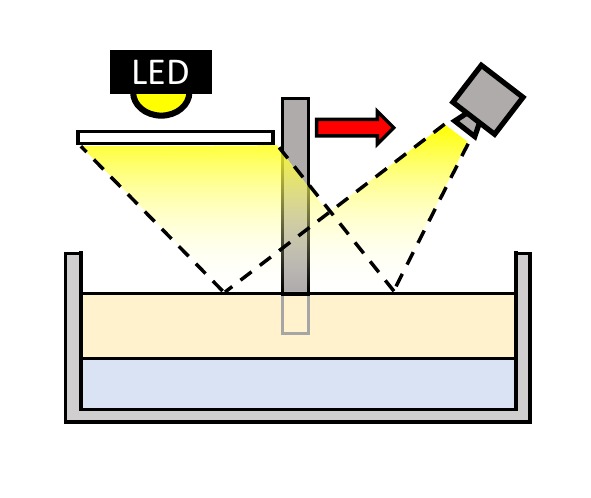}}\\\vspace{0.5em}
  \subcaptionbox{\label{fig:SETUP:TOP}}{\includegraphics[width=0.67\columnwidth]{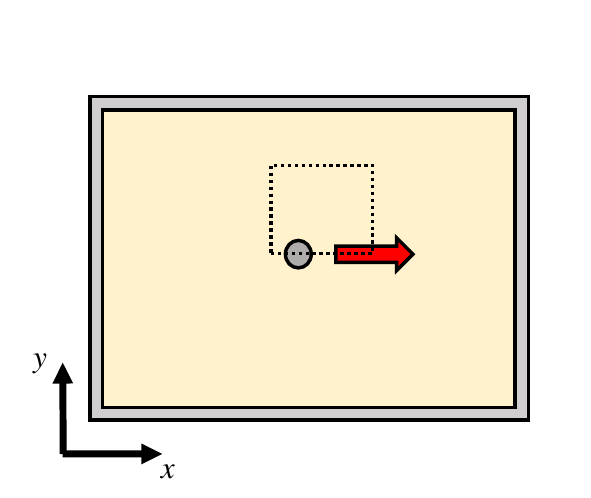}}
  \caption{Schematic of the experimental set-up. (a) PIV set-up with two cameras, (b) texture set-up with the backlit acrylic plate and (c) birds-eye-view of the suspension surface with the dotted square indicating the investigation region of the present study. In the images, the beige layer represents the suspension and the blue layer beneath it represents a layer of Fluorinert.}
  \label{fig:SETUP} 
\end{figure}

The experimental set-ups are shown in figure~\ref{fig:SETUP}. Both experiments are conducted in a $1$~m~$\times 0.5$~m tank. The tank is first filled with a $15$~mm layer of high density, low viscosity Fluorinert oil (FC74) \citep{Loimer2002, Peters2014,Han2018,Romcke2021a}, followed by a $15$~mm thick suspension layer consisting of cornstarch (\textit{Maizena maisstivelse}) and a sucrose-water solution ($50$\%~wt) at a nominal volume fraction of $\phi=0.36$ \citep{Romcke2021a}, defined as
\begin{equation}
\label{eq:pf}
    \phi=\frac{(1-\beta)m_s/\rho_s}{(1-\beta)m_s/\rho_s + m_l/\rho_l + \beta m_s/\rho_w}.
\end{equation}
Here, $\beta=11$~\% is the water content in the starch, while $m_s$ and $m_l$ are the measured mass of starch and sucrose solution, respectively. The densities of the starch, sucrose solution and water are $\rho_s=1.63$~g/ml, $\rho_l=1.23$~g/ml and $\rho_w=1.0$~g/ml, respectively. We mix the suspension for two hours before it is loaded into the tank. The suspension floats atop the denser Fluorinert ($\sim 1.8$~g/ml), which ensures a near stress free bottom boundary and makes the system approximately 2D \citep{Peters2014}. A $25$~mm diameter ($D$) cylinder is submerged in the suspension and is traversed at a velocity of $U_c=0.14$~m/s; the effect of changing $U_c$ is the subject of a previous study \citep{Romcke2021a}. Both the cylinder velocity ($U_c>0.06$~m/s) and volume fraction ($\phi_m<\phi<\phi_0$) are in a range where dynamic jamming is known to occur for this suspension \citep{Romcke2021a}. 

The suspension is pre-sheared by towing the cylinder back and forth equivalent to an actual run, before any measurements are taken. When capturing particle images for the PIV, two 4 megapixel high-speed cameras (Photron FASTCAM Mini WX100) view the suspension surface in front and behind the traversing cylinders (figure~\ref{fig:SETUP:PIV}). Pulsed LED lighting was used to illuminate the surface and was synchronised with the camera acquisition at 750 Hz. The particle images were converted to velocity fields with LaVision DaVis 8.4.0 PIV software. An initial pass was performed with $96$~pixels~$\times$~$96$~pixels square interrogation windows, followed by two passes with circular interrogation windows with decreasing size ending at $48$~pixels~$\times$~$48$~pixels. For all passes, the interrogation windows have a $50$\% overlap. The resulting instantaneous velocity fields are stitched together in post processing, masking out the cylinder in each frame. This results in a velocity field fully surrounding the cylinder.

As mentioned above, only minor adjustments to the set-up are needed in order to observe the surface features. This is illustrated in figure~\ref{fig:SETUP:TEX}. Here, a single camera is positioned such that it views a direct reflection of a backlit, semi-transparent, acrylic sheet on the suspension surface. This enhances any surface features not captured by the PIV; note that the tracer particles used for PIV also interfere with the detection of the surface topology, which is why a separate campaign was used for surface texture measurements. Figure~\ref{fig:SETUP:TOP} gives a birds eye view of the suspension surface. For the scope of this work, we focus on the region indicated by the dotted square. See \citet{Romcke2021a} for details on the full velocity field.

For the texture images, LaVision Davis 8.4.0 was used to find a third order calibration polynomial, mapping the image coordinates ($i, j$) to the lab coordinates ($x, y$). Matlab was used for all further processing of the texture images. Here, for consistency, the results are always plotted in the calibrated lab coordinate system ($x, y$). However, calculations on the surface features are done in pixel coordinates ($i,j$), and the results in ($i, j$) are mapped to ($x, y$) with the calibration polynomial.

\begin{figure*}
    \centering
    \includegraphics[width=0.9\textwidth]{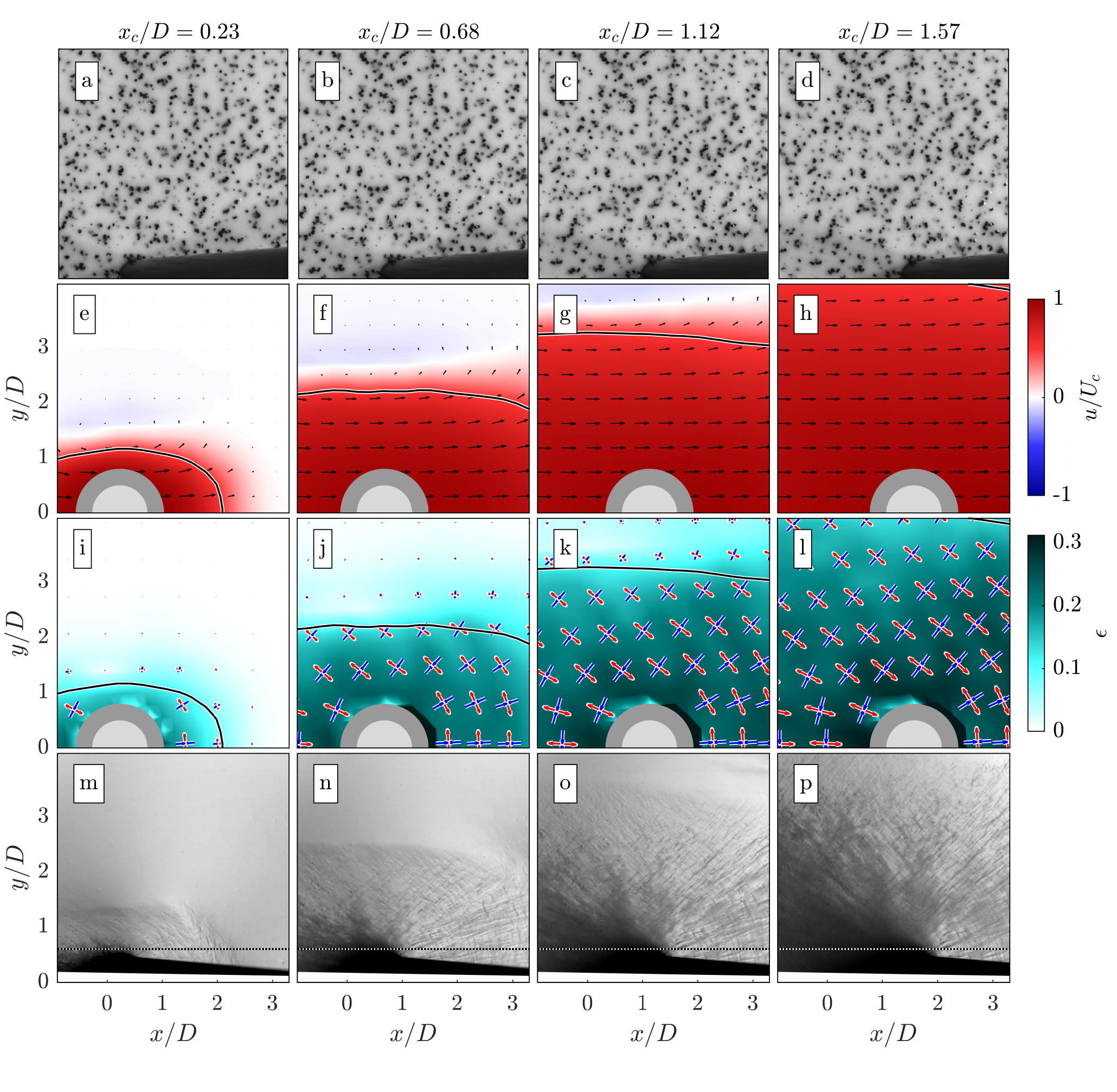}
    \caption{Time series comparison between PIV and texture images. $x_c$ represents the location of the cylinder. (a-d) Raw PIV particle images. (e-h) Resulting PIV velocity fields with superimposed velocity vectors. (i-l) Accumulated strain ($\epsilon$) with superimposed eigenvectors. Red represents the direction of stretch ($\mathbf{n}_1$), while blue indicates the direction of compression ($\mathbf{n}_2$). For clarity, $\mathbf{n_i}$ are scaled with $\epsilon$. In (e-l) the jamming front ($0.5U_c$ contour) is represented by the black line. (m-p) Surface features as the front propagates. In order to avoid the cylinder shadow, the analysis is restricted to the region above the dotted line.} 
    \label{fig:DATA} 
\end{figure*}

Examples of both the PIV and texture images are shown in figure~\ref{fig:DATA}. The cylinder position is denoted $x_c$, which starts as $x=0$ and moves in the positive x-direction. The raw particle images (figures~\ref{fig:DATA}a-d) do not provide any information on the suspension texture. The resulting velocity fields from the PIV analysis of these images are shown in figures~\ref{fig:DATA}e-h. Note the sharp transition in velocity that propagates away from the cylinder as it moves through the flow. By the end of an experimental run, the whole field of view is moving with the cylinder as shown in figure~\ref{fig:DATA}h. The $0.5U_c$ contour is represented by the black line and used as a proxy for the position of the jamming front as is common in previous studies \citep{Waitukaitis2012, Peters2014, Peters2016, Han2016, Han2018, Han2019, Han2019b,Romcke2021a}. 

From the velocity data, we estimate the local accumulated strain. The strain is shown to be an important parameter with regards to jamming. Given that the suspension is subjected to a sufficient amount of stress, an intrinsic onset strain dictates the amount of strain needed before the suspension transitions into a jammed state \citep{Majumdar2017, Han2016, Han2019b, Romcke2021a}. The nominal value of the onset strain depends on the volume fraction \citep{Han2016, Han2018, Han2019b}. We define the strain in the same manner as \cite{Romcke2021a}. In short, by estimating the movement of the material points $\mathbf{x_p}(\mathbf{X},t)=\mathbf{X}+\int_0^t\mathbf{u}(\mathbf{x_p}(\tau),\tau)d\tau$, we calculate the deformation gradient tensor $\mathbf{F}=\frac{\partial\mathbf{x_p}}{\partial\mathbf{X}}$, where $\mathbf{X}$ is the position of the material points at $t=0$. From the deformation gradient the left stretch tensor $\mathbf{V}$ is acquired from a polar decomposition $\mathbf{F}=\mathbf{VR}$. The tensor $\mathbf{V}$ has eigenvalues and eigenvectors, $\lambda$ and $\mathbf{n}$, respectively. Here, we employ the Eulerian logarithmic strain tensor \citep{Nasser2004}
\begin{equation}
    \mathbf{e}=\sum_i \ln(\lambda_i) \mathbf{n}_i\otimes\mathbf{n}_i.
    \label{eq:Strain}
\end{equation} 
Eigenvalues are ordered $\lambda_1>\lambda_2$, such that $\mathbf{n}_1$ and $\mathbf{n}_2$ signify the direction of stretch and compression, respectively; $\mathbf{n}_1$ and $\mathbf{n}_2$ are orthogonal. Figures~\ref{fig:DATA}i-l show the evolution of the norm of the strain tensor $\epsilon=||\mathbf{e}||$. The strain ($\epsilon$) at the jamming front is relatively constant throughout an experiment \citep{Romcke2021a}, and measured to be $\sim 0.14$ in the case presented here. In the current work, we will focus on the orientation of the eigenvectors. Though the strain field is relatively homogeneous by the end of an experimental run (figure~\ref{fig:DATA}l), the superimposed eigenvectors show that the direction in which the suspension is stretched and compressed depends on the location in the flow.

Finally, a time series of the texture experiment is presented in figure~\ref{fig:DATA}m-p. Due to the specific lighting conditions explained above, this experiment reveals features not visible in the PIV particle images. Two key observations form the scope of this work. First, like the velocity field, there is a sharp transition between the regions with and without surface structures. This transition propagates from the cylinder, into the suspension, leaving a textured surface in its wake. We observe a change from a reflective to a matte surface indicative of a dilated suspension \citep{Brown2012, Maharjan2021, White2010, Smith2010}. Second, by comparing the  eigenvectors of the strain tensor (figure~\ref{fig:DATA}i-l) and the orientation of the surface features (figure~\ref{fig:DATA}m-p), there appears to be a connection. More specifically, the eigenvectors and surface features appear to be oriented in the same direction at the same locations in the flow, suggesting the raw surface images may hold quantifiable information akin to the PIV. This is explored further below.


\section{Analysis methodology}
\label{sec:Method}
The aim of this study is to quantify the structures observed at the free surface of a dense suspension undergoing dynamic jamming. In this section, we present a method that is able to identify surface features and their orientation.  In short, images of the free surface are divided into interrogation windows, and the 2D fast Fourier transform (FFT) of the local windows is used as a basis for quantifying these structures, which is presented in section~\ref{sec:Method:FFTSecAvg}. Section~\ref{sec:Method:OptParam} establishes a basis by which this process can be optimised and determines the optimal parameter values used in the remainder of this work.

\subsection{FFT and sector averaging}
\label{sec:Method:FFTSecAvg}
\begin{figure}
    \centering
    \includegraphics[width=\columnwidth]{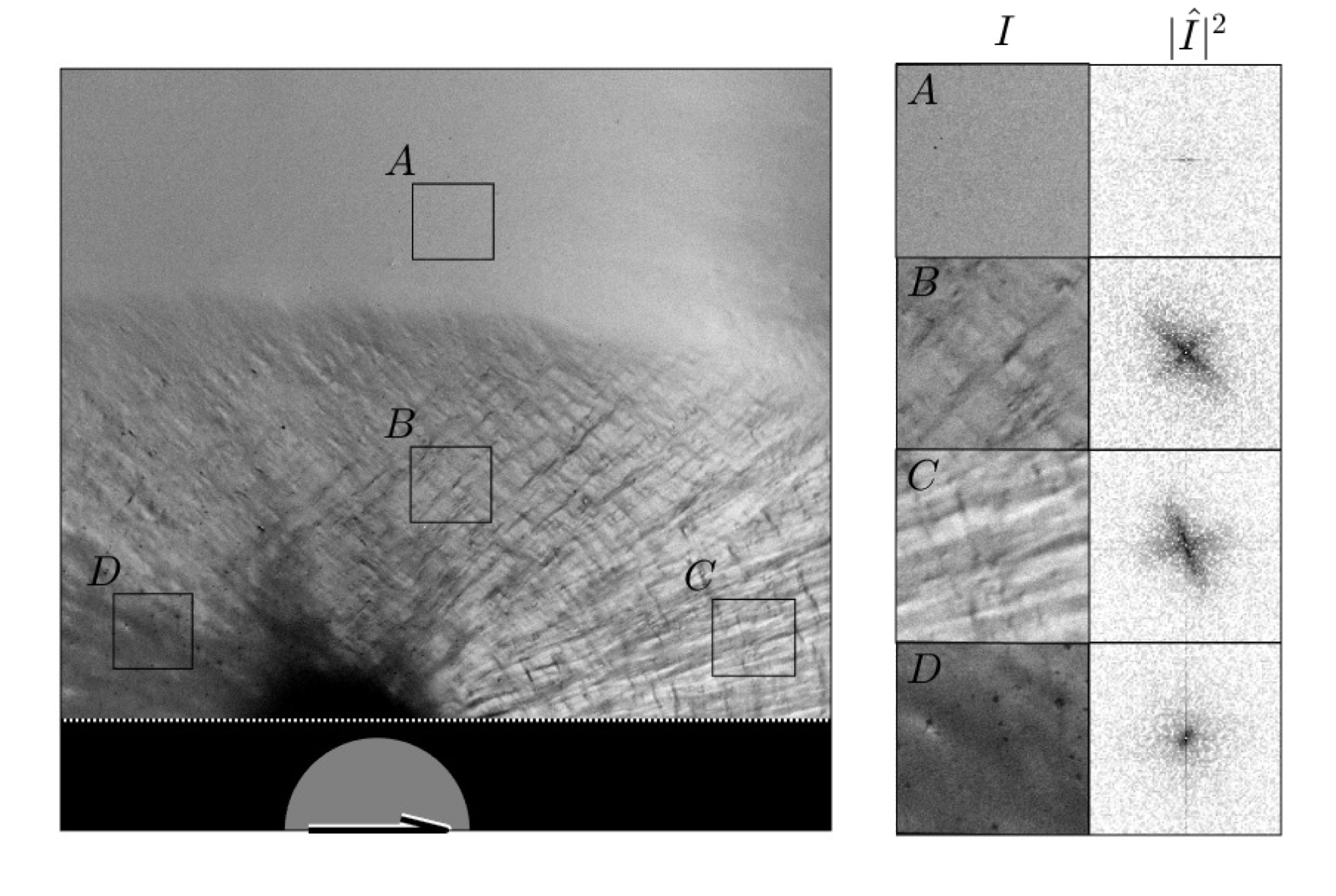}
    \caption{A snapshot of the suspension surface. Example interrogation windows are presented in the right columns with both the pixel intensity ($I$) and the 2D power spectra ($|\hat{I}|^2$). The intensity scale of the power spectra is plotted logarithmically. $200\times200$ pixel interrogation windows representing (A) no features, (B and C) distinct features, (D) weak features.} 
    \label{fig:FOCUS} 
\end{figure}

Here, in order to extract local information from the texture images, such as how dominant the features are and what orientation they have, we divide the frame into interrogation windows. Figure~\ref{fig:FOCUS} gives examples of four representative windows, which we will focus on in this section. As a first step towards quantifying the surface features, we take a 2D~FFT of each window ($I(i, j)\xrightarrow{\mathcal{F}}\hat{I}(k_i, k_j)$); using an FFT to gain information on the surface texture is a common practice for interfacial flows, e.g., \citep{Zhang1996, Loimer2002, Timberlake2005, Singh2006}. The mean pixel intensity of the interrogation window is subtracted before calculating the FFT. Here, $I$ and $\hat{I}$ represent the intensity in the image and wave number domain, respectively. 

The resulting FFTs of the interrogation windows are seen in the right column of figure~\ref{fig:FOCUS} represented here by the power spectral density $|\hat{I}|^2$. Notice how the streaks in the interrogation windows are reflected in the corresponding power spectra. For B and C, the power spectra show clear features orthogonal to the streaks observed in the image. This trend is also observed for D, though clustered at lower wave numbers. Window A, however, has an almost perfectly homogeneous intensity, which is reflected in the power spectrum by predominantly exhibiting values at the noise level. An unwanted feature is also revealed by the FFT. As seen in the power spectra of A and D, a signal is observed along the lines $k_i=0$ and $k_j=0$. Though the background subtraction removes most of these features, they are still significant for low signal-to-noise ratio regions, e.g., A and D. For any further processing, this issue is addressed by masking out the values at the lines $k_i=0$ and $k_j=0$. 

\begin{figure}
  \centering
  \subcaptionbox{\label{fig:Schem:SA}}{\includegraphics[width=0.38\columnwidth]{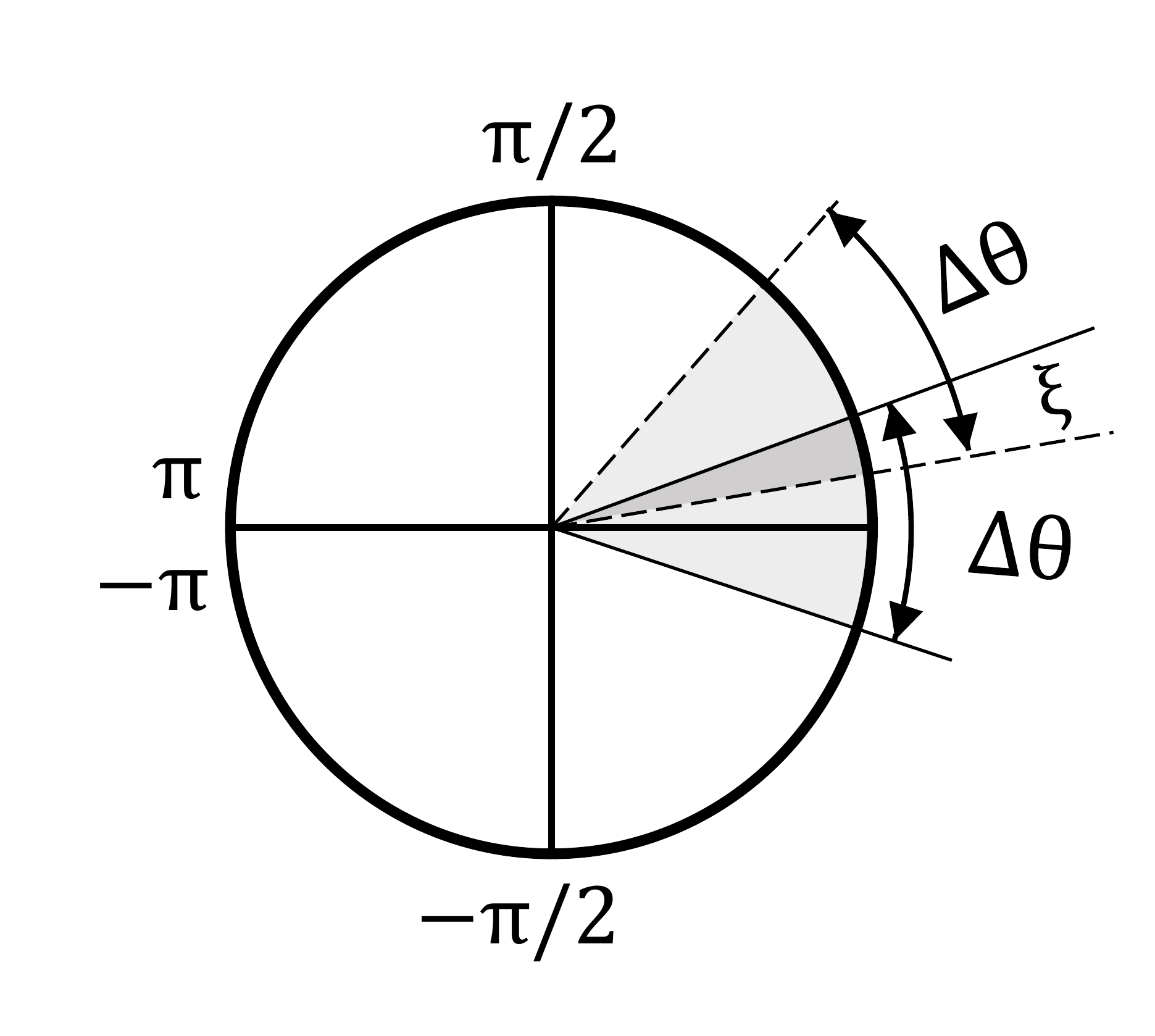}}
  \subcaptionbox{\label{fig:Schem:PS}}{\includegraphics[width=0.27\columnwidth]{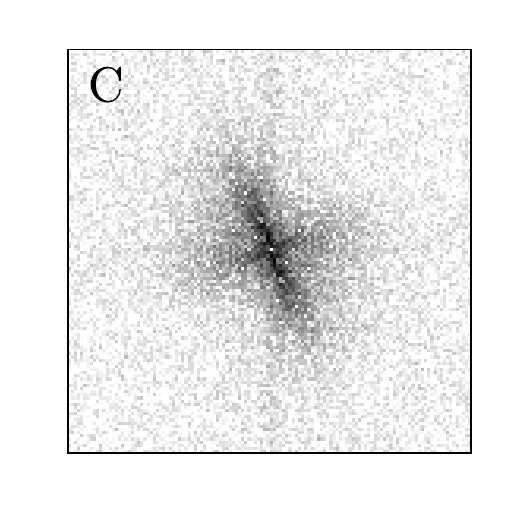}}
  \subcaptionbox{\label{fig:Schem:RES}}{\includegraphics[width=0.3\columnwidth]{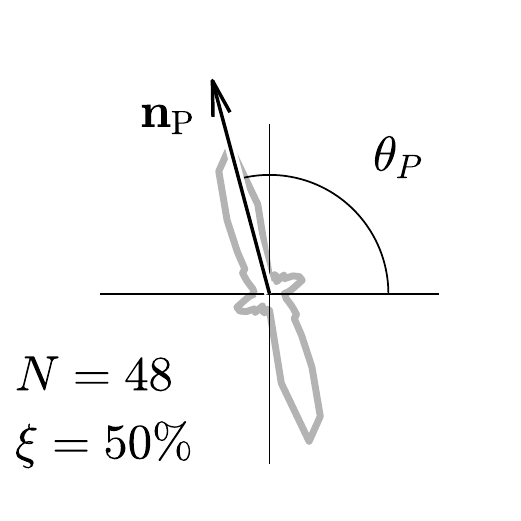}}
  \caption{Schematics of the sector averaging. (a) Two neighbouring sectors with overlap. (b) Power spectra of interrogation window C with (c) resulting sector average curve ($f$) calculated with $N=48$ sectors and $\xi=50$\% overlap. The vector $\mathbf{n}_P$ and the angle $\theta_P$ represent the orientation of the peak.}
  \label{fig:SectorAverage} 
\end{figure}

The general shape of the power spectra encodes both how dominant the features are and in what direction they are oriented. We extract the shape of a power spectrum by taking a sectional average. First, we denote the wave vector $\mathbf{k}=(k_i,k_j)$, so that in polar coordinates, $k=|\mathbf{k}|$ with the angle $\theta=\tan^{-1}(k_j/k_i)$. The averaging procedure is taken over the wave number $k$ ranging from $0$ to $0.5$~pixel$^{-1}$ in sectors of size $\Delta \theta$. In polar coordinates, the average over a sector is represented by the integral
\begin{equation}
    f(\theta, \Delta \theta) = \frac{8}{\Delta\theta}\int_{0}^{1/2} \int_{\theta-\Delta\theta/2}^{\theta+\Delta\theta/2} |\hat{I}|^2(k, \theta)k\, \text{d}\theta\,\text{d} k.
    \label{eq:SecAvgCurve}
\end{equation}
Numerically, given the discrete values of $|\hat{I}|^2$, the circle is divided into a number ($N$) of equally spaced angular sectors with a relative overlap $\xi$, shown schematically in figure \ref{fig:Schem:SA}. Note that $\Delta \theta = 2\pi/(N(1-\xi))$. The sector average is calculated as the mean of $|\hat{I}|^2$ contained within the sector. Figures \ref{fig:Schem:PS} and \ref{fig:Schem:RES} show how the trend in the power spectra is reflected in the resulting curve $f$. 

We use the sector average curve, $f$, to extract key features of a power spectrum. First, we calculate the shape factor, $S$, of the sector average curve in order to distinguish between an interrogation window with and without surface structure. $S$ is a measure of how much a shape resembles a circle. The shape factor is defined such that it compares the perimeter ($\delta$) of $f$ with the circumference of a circle with the same area ($\alpha$) as $f$, i.e.,  
\begin{equation}
    S=\frac{2\sqrt{\pi}\sqrt{\alpha}}{\delta}.
    \label{eq:ShapeFactor}
\end{equation}
The shape factor $S$ is in the range $0$ to $1$, where $1$ represents a perfect circle. The power spectrum of window A from figure~\ref{fig:FOCUS} is relatively uniform, and is expected to show values of $S$ close to unity. This can also be seen by substituting in a constant value for $|\hat{I}|^2$ in \eqref{eq:ShapeFactor} resulting in $f(\theta)=$~constant. Window C, on the other hand, is expected to show values distinguishable from a perfect circle as we observe clear angular dependencies.  Secondly, we take the orientation of the peak of $f$ to represent the orientation of the dominant surface features. In the image coordinates $(i,j)$, we denote the orientation of the peak $\theta_P^*$ which defines the unit vector $\mathbf{n}_P^*=(\cos{\theta_P^*}, \  \sin{\theta_P^*})$. The unit vector is mapped to the spatial coordinate system with the calibration, such that $\mathbf{n}_P=(\cos(\theta_P), \ \sin(\theta_P))$, where $\theta_P$ represents the orientation of the peak in the $(x,y)$-system. Note that the Fourier transform is symmetric about the origin, thus two equal sized peaks separated by an angle $\pi$ are observed in figure \ref{fig:Schem:RES}. As a representation of the main direction of the structures, we only focus on peaks in the upper half plane $0\leq \theta_P<\pi$.


\subsection{Determining processing parameters}
\label{sec:Method:OptParam}
Some parameters have been presented in the previous section that affect the sector average curve $f$. Here, we focus on the size of the interrogation window, as well as the number of sectors ($N$) and the sector overlap ($\xi$). Adjusting these parameters will consequently have an impact on $S$ and $\theta_P$. As one of our aims is to be able to distinguish between regions with surface features (such as window C) from regions without surface features (such as window A), we choose to use a set of processing parameters that will maximize the difference in $S$ between these two scenarios. 

\begin{figure*}
  \centering
  \subcaptionbox{\label{fig:secavgA}}{\includegraphics[width=\columnwidth]{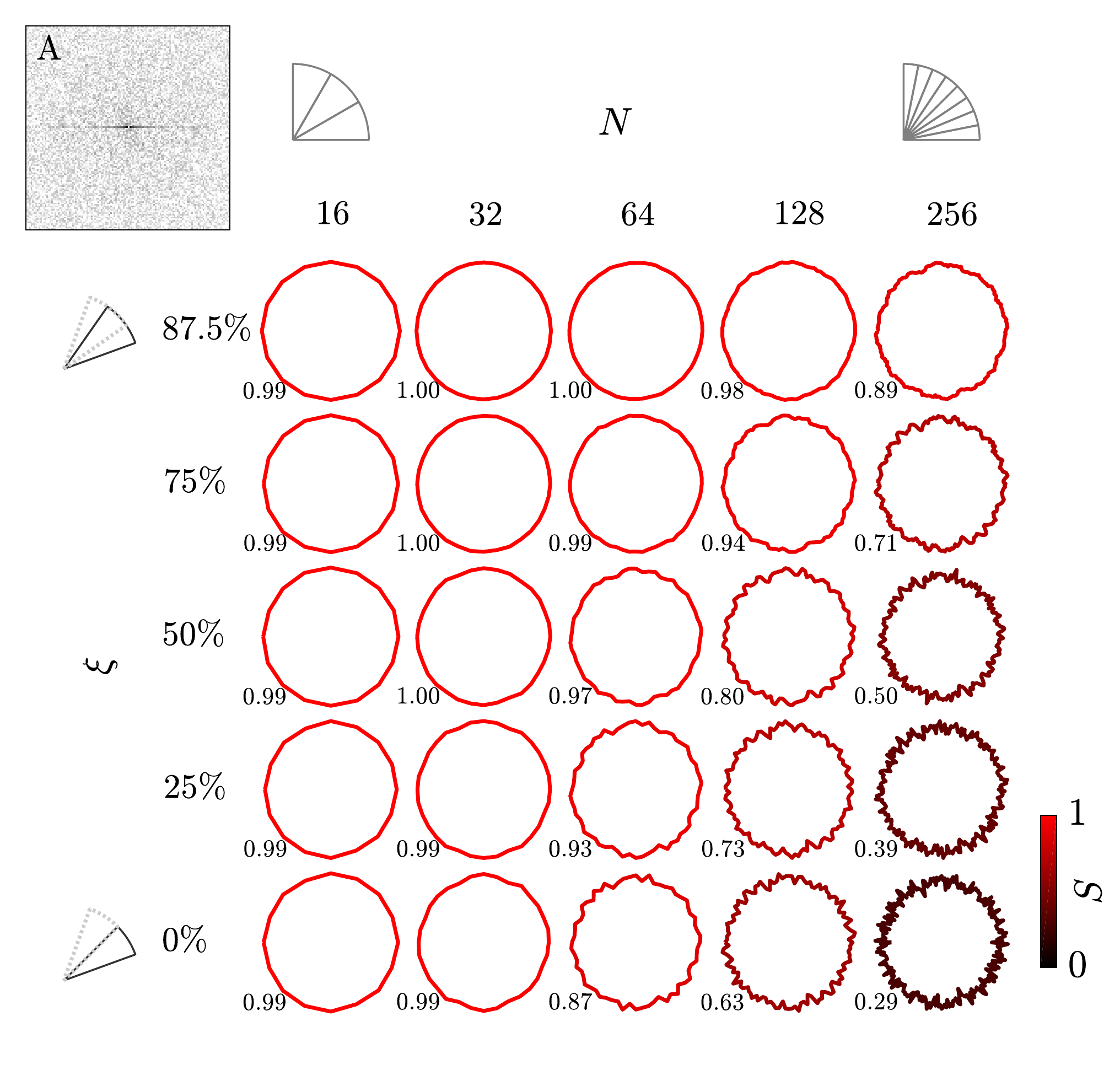}}
  \subcaptionbox{\label{fig:secavgC}}{\includegraphics[width=\columnwidth]{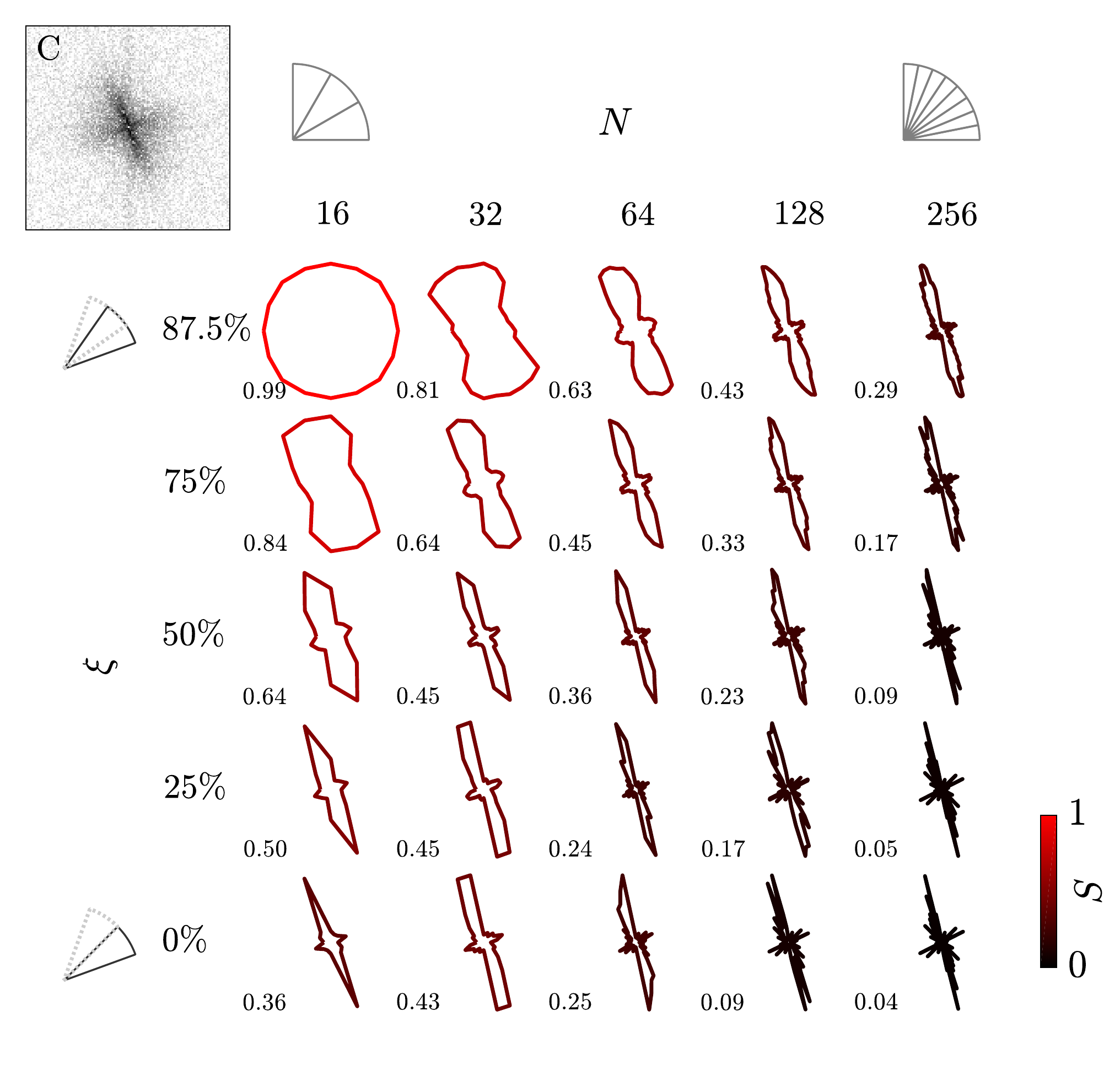}}
  \caption{Sector average curves ($f$) for $256\times256$ pixels interrogation windows of (a) A and (b) C over a range of overlap ($\xi$) and number of sectors ($N$). The resulting shape factor ($S$) is indicated by the color scheme and also presented numerically for each curve. For the remainder of this work, we choose $N=128$ and $\xi=75$~\%.}
  \label{fig:SectorAverageRes} 
\end{figure*}

Trends of $S$ for window A and C over a range of processing parameters ($N$ and $\xi$) are presented in figure \ref{fig:SectorAverageRes}. As expected, the sector average curve, $f$, for window A (figure \ref{fig:secavgA}) takes on a circular shape, while for window C (figure~\ref{fig:secavgC}), the curve indicates clear peaks orthogonal to the streaks on the surface. This is also reflected by the values of $S$ showing higher values for window A compared to C. However, in the extreme cases, the difference in $S$ tends to be small. For example, choosing a large number of sectors and a small overlap, even though the the resulting sector average curve from C shows clear peaks and a low shape factor, the shape factor of A is also reduced. On the other hand, choosing few sectors with a large overlap yields an almost perfect circular result for window A, but the peaks in C are no longer clear and the shape factor is higher. Figure \ref{fig:SectorAverageRes} indicates that there is an optimal combination of $N$ and $\xi$ that would maximize the difference in shape factor $S$, such that the shape factor yields a clear distinction between the two scenarios. 

We point out that the result presented in figure \ref{fig:SectorAverageRes} only compares two locations in the flow (A and C) for one window size ($256 \times 256$~pixels) from a specific snapshot of the flow. However, the trend is the same when comparing window A with window B and D over a range of window sizes from~$64 \times 64$~to~$512 \times 512$~pixels. After considering different window sizes and values for $N$ and $\xi$, the overall difference in $S$ appears to converge with increasing window size at $256 \times 256$~pixels. Setting $N=128$ and $\xi=75$~\% gives a sufficiently large difference in $S$, while simultaneously ensuring a satisfactory angular resolution for establishing $\theta_P$. For the remainder of this work, we will use a window size of $256 \times 256$~pixels, with $N=128$ and $\xi=75$~\% when computing the sector average curve, $f$. It would be important to note that the ``optimal" is dependent on the specific experimental set-up and that the results found here would not be universally optimal akin to how PIV processing parameters are optimised individually for each experiment.


\section{Results}
\label{sec:Results}
The method outlined in section \ref{sec:Method} is now applied to the full texture images. Section \ref{sec:Result:FULLFIELD} presents results from the texture images alone. By setting a threshold value for $S$, we identify the textured region and in section \ref{sec:Result:TEXPROP} we give an estimate of how fast this region propagates into the suspension. In section \ref{sec:Result:TEXCORR}, the data extracted from analysing the texture images are directly compared to the PIV data. First we present the combined evolution of the shape factor and velocity field. We then show that in certain regions of the flow, the direction of the texture and the eigenvectors of the strain tensor are predominantly oriented in the same direction. 

\subsection{Quantifying the texture for the full field}
\label{sec:Result:FULLFIELD} 

\begin{figure*}
  \centering
  \subcaptionbox{\label{fig:FULL:CF}}{\includegraphics[width=\textwidth]{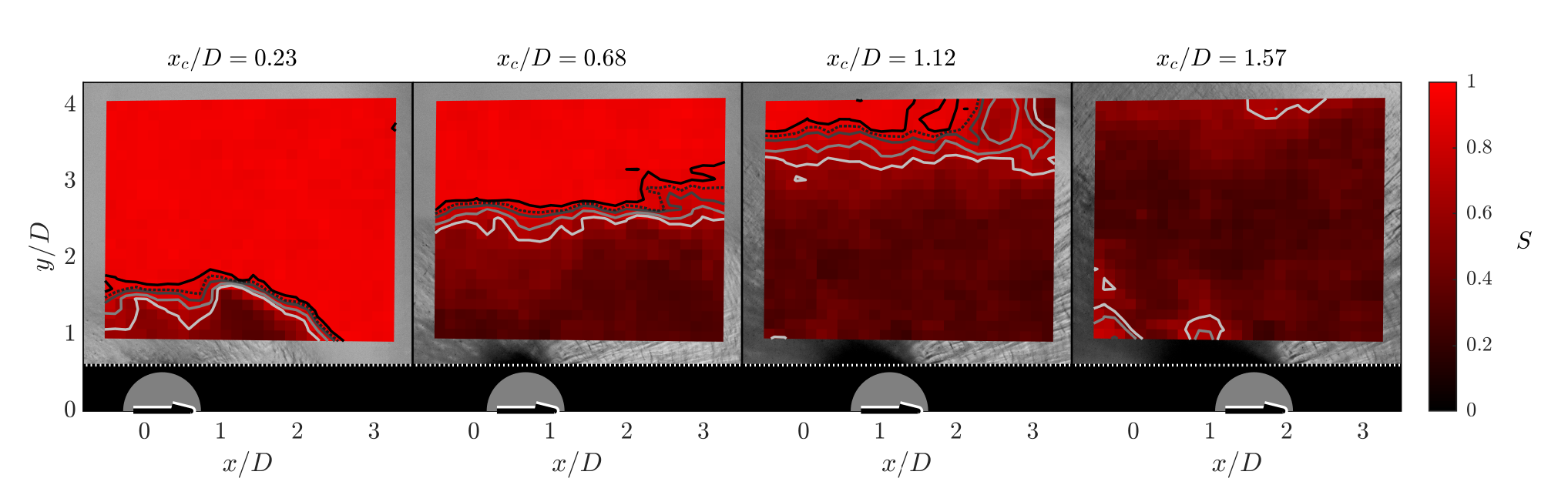}}\\
  \subcaptionbox{\label{fig:FULL:Q}}{\includegraphics[width=\textwidth]{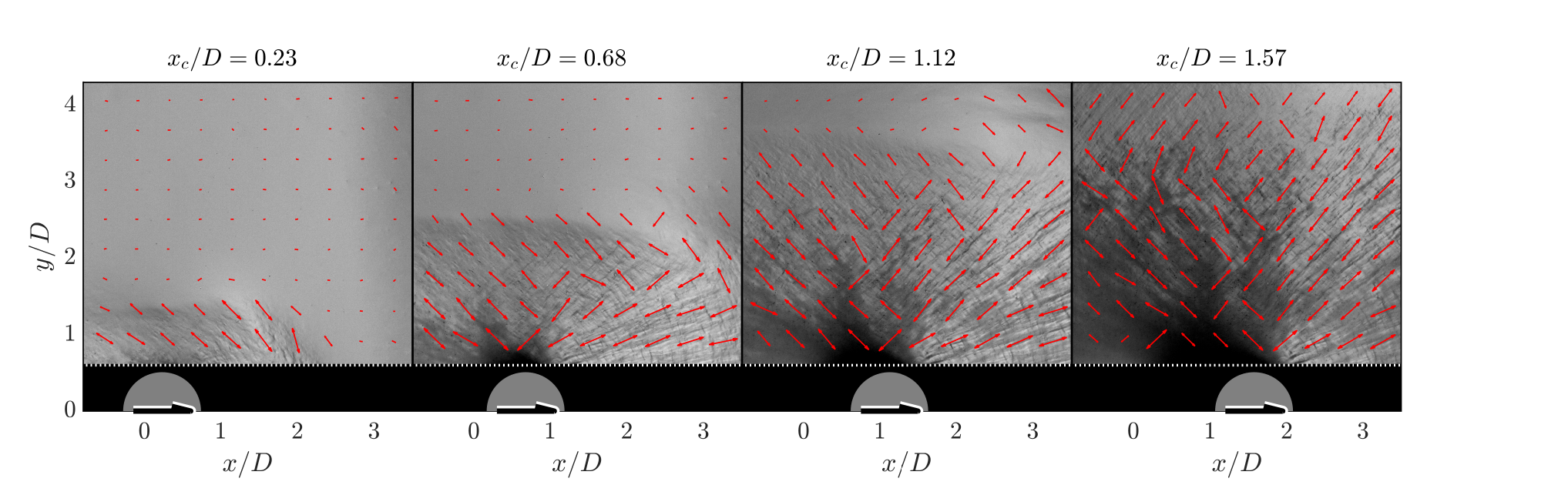}}
  \caption{Texture analysis on the full field of the snapshots in figure \ref{fig:DATA}m-p.  (a) Shape factor $S$. The contours represent $S=0.6$, $0.7$, $0.8$ $0.85$ and $0.9$, where the $0.85$ is represented by the dotted line. (b) Orientation of the peak ($\mathbf{n}_P$) from the sector average, which is represented by the vectors superimposed on the corresponding surface image. Every third vector is plotted to reduce clutter, and scaled with $1-S$. In addition, $\mathbf{n}_P$ is rotated by $\pi/2$ such that the vectors point in the same direction as the surface streaks rather than normal to them.}
  \label{fig:FULL} 
\end{figure*}

From the method presented above, we are capable of identifying if the surface shows features ($S$), and in what direction the features are oriented ($\mathbf{n}_P$). Both $S$ and $\mathbf{n}_P$, are presented in figure~\ref{fig:FULL}. The analysis is conducted on the full field of the time series shown in figure~\ref{fig:DATA}m-p, with the processing parameters found in section~\ref{sec:Method:OptParam}. In addition, we let neighbouring interrogation windows overlap with $75$\%. 

Figure~\ref{fig:FULL:CF} shows that the shape factor clearly separates the flow into two regions. Five iso-contours, $S=0.6$, $0.7$, $0.8$, $0.85$ and $S=0.9$, are superimposed on the $S$-field. Here, these contours tend to cluster at the transition between the textured and texture-free surface. Analogous to the $0.5U_c$ contour being used to identify the position of the jamming front from PIV data, a contour level between $S=0.6$ and $S=0.9$ identifies the position of the front from the texture data. In the later stages of an experimental run, dilation renders the surface matte. Due to the increase in $S$ observed in figure~\ref{fig:FULL:CF}, it becomes increasingly difficult to identify the preferred texture orientation in some parts of the flow. This is most noticeable in the wake of the cylinder.

The corresponding orientation of the surface features are indicated in figure \ref{fig:FULL:Q}. As noted in figures \ref{fig:SectorAverage} and \ref{fig:SectorAverageRes}, the peak in $f$ is oriented normal to the direction of the dominant streaks in the texture image. For clarity in figure~\ref{fig:FULL:Q}, the vectors indicating the peak angle, $\mathbf{n}_P$, are rotated by $\pi/2$ such that they are oriented parallel to the surface streaks. Here, we plot both $\mathbf{n}_P$ and $-\mathbf{n}_P$, representing the symmetry of the power spectra. 

The orientation of the vectors can be directly compared to the actual texture image. Notice that fore (aft) of the cylinder, the vectors generally tend to be forward (backward) leaning, reflecting the dominant streaks in the region. The region roughly around the same $x$-location as the cylinder, exhibits a crosshatch pattern \citep{Chang1990, Albrecht1995}. This is more clearly indicated in figure \ref{fig:FOCUS} by window B. As we only report the most dominant peak in $f$, the vectors $\mathbf{n}_P$ generally show a mix of forward-leaning and backward-leaning in this region. Notice the similarity with the eigenvectors plotted in figures~\ref{fig:DATA}i-l; $\mathbf{n}_1$, representing direction of stretch, is backwards-leaning, while $\mathbf{n}_2$, representing compression, forward leaning. The similarity between $\mathbf{n}_P$ and the strain eigenvectors will be addressed in greater detail in section \ref{sec:Result:TEXCORR}.

\subsection{Propagation of the texture transition}
\label{sec:Result:TEXPROP}

\begin{figure}
  \centering
  \subcaptionbox{\label{fig:TEXFRONTPROP:PROF}}{\includegraphics[width=0.9\columnwidth]{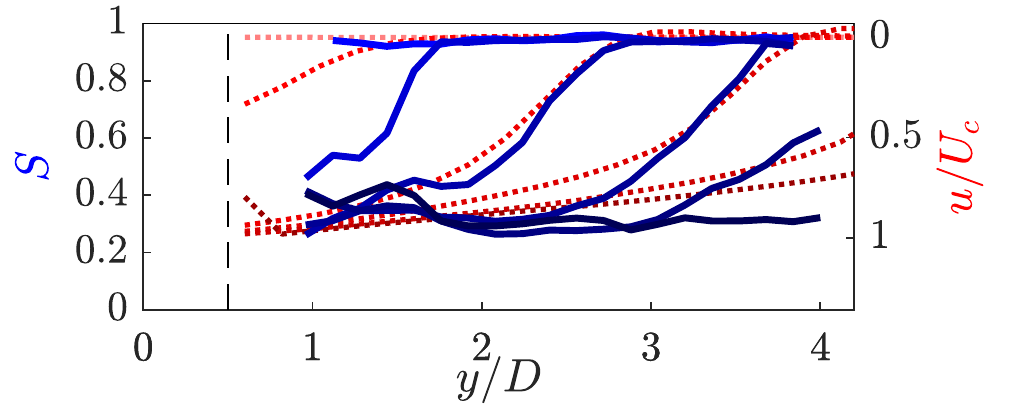}}\\
  \subcaptionbox{\label{fig:TEXFRONTPROP:POS}}{\includegraphics[width=0.9\columnwidth]{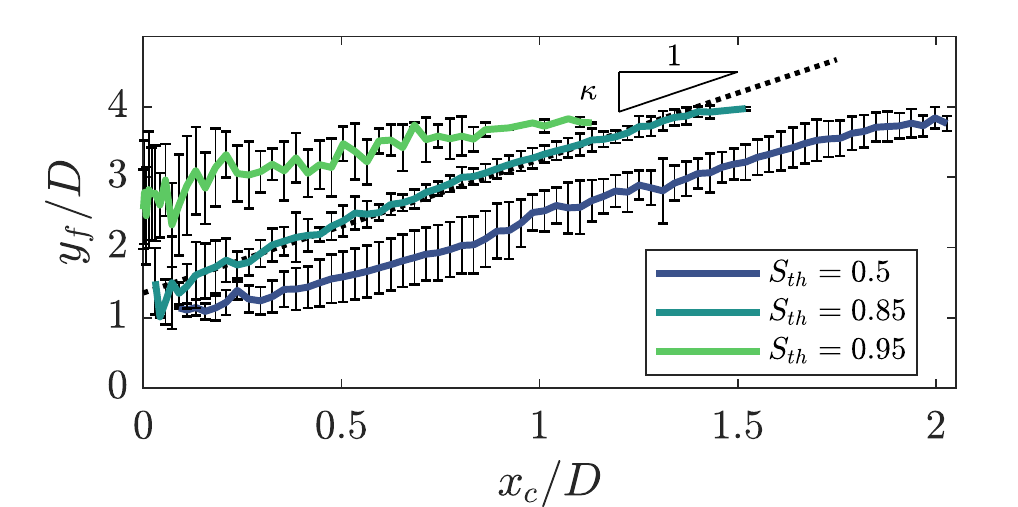}}\\
  \subcaptionbox{\label{fig:TEXFRONTPROP:LF}}{\includegraphics[width=0.9\columnwidth]{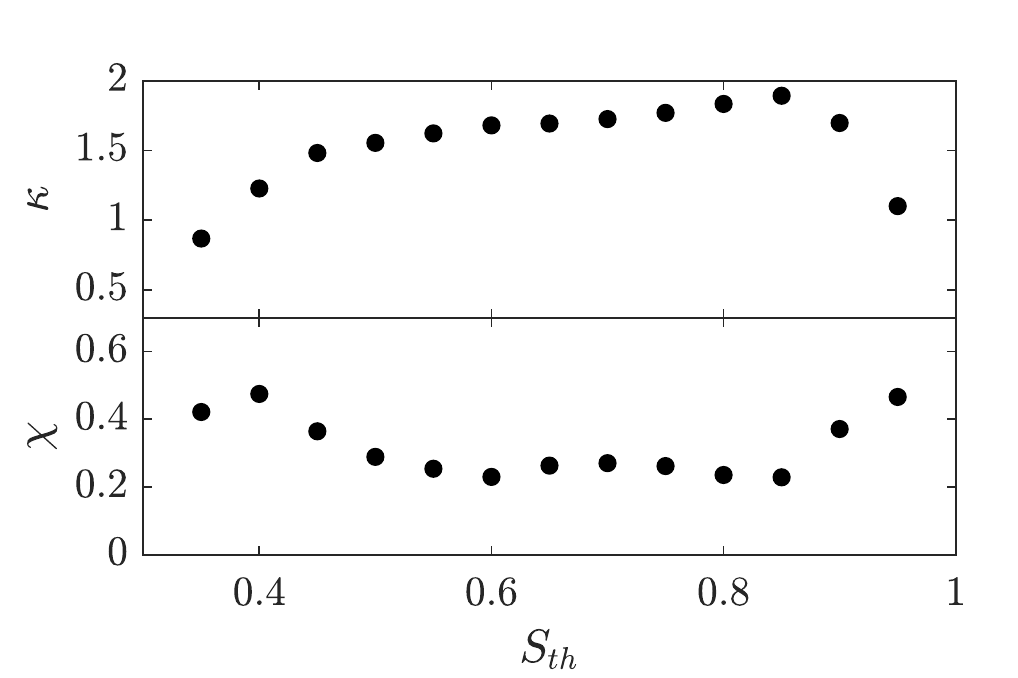}}
  \caption{(a) Profiles of the shape factor (solid blue lines) and velocity (dotted red lines) for an experimental run in vertical cross sections taken at the location of the cylinder ($x=x_c$). The black dashed line indicates the location of the cylinder surface. The color intensity indicates different $x_c$ as the cylinder moves through the flow. Velocity profiles are flipped to emphasize the similarity between $S$ and $u$ (b) Position of the front, $y_f$, for three different threshold values of the shape factor. The slope of the linear regression line is denoted $\kappa$. (c) Slope, $\kappa$, and RMS of the error of the regression line, $\chi$, as functions of the threshold value, $S_{th}$.}
  \label{fig:TEXFRONTPROP} 
\end{figure}

Figure \ref{fig:TEXFRONTPROP} establishes the location of the texture transition, and estimates its propagation velocity into the suspension. Here, we focus on the transverse direction relative to the cylinder velocity. As with the jamming front position being identified by the $0.5U_c$ contour, we will identify the texture transition by a contour value in $S$. We will compare how fast the jamming front and texture transition propagates into the suspension.

The front propagation factor, denoted here by $\kappa$, is defined as the relation between the speed of the jamming front and the speed of the perturbing body. In the transverse direction relative to the cylinder velocity, the front propagation factor is the time derivative of the jamming front's $y$-position ($y_f$) relative to the speed of the cylinder ($U_c$). Since 
\begin{equation}
    \kappa=\frac{dy_f/dt}{U_c}\approx \frac{\Delta y_f/\Delta t}{\Delta x_c/\Delta t}=\frac{\Delta y_f}{\Delta x_c}, 
    \label{eq:FrontProp}
\end{equation}
rather than estimating a time derivative, we will focus on the equivalent relation $\Delta y_f/\Delta x_c$. As such, for the texture data, we start by establishing the position of the texture transition.

As seen by the superimposed contours in figure \ref{fig:FULL:CF}, the position of the texture transition depends on the choice of contour level. Figure \ref{fig:TEXFRONTPROP:PROF} shows vertical cross sections of the shape factor taken as the cylinder translate in the $x$-direction ($S(x=x_c,y)$, where $x_c$ represents the instantaneous cylinder position). As with the velocity profiles (indicated by the dotted lines in figure \ref{fig:TEXFRONTPROP:PROF}), $S$ is not a perfect step, and as pointed out above, the position of the front will depend on the choice of the contour level. We use $y_f$ to denote the position of the texture transition and is given implicitly from the shape factor profiles as $S_{th}=S(x=x_c,y_f)$. Here, $S_{th}$ represents the contour level, or the threshold value for $S$ separating textured from texture free surface. Numerically, the front position is acquired by linearly interpolating the shape factor profiles, e.g. figure \ref{fig:TEXFRONTPROP:PROF}. 

The resulting front position ($y_f$) from all experimental runs is plotted in figure \ref{fig:TEXFRONTPROP:POS} for three different values of $S_{th}$.  Due to variation in the front position shown in figure \ref{fig:TEXFRONTPROP:POS}, calculating the relation $\Delta y_f/\Delta x_c$ directly has large uncertainties associated with it. Instead, we report the slope of the linear regression line through the data, with the root mean square of the error denoted as $\chi$. Figure \ref{fig:TEXFRONTPROP:LF} shows $\kappa$ and $\chi$ as functions of $S_{th}$. The error tends to show a minimum in the range $0.5 < S_{th} < 0.9$ with values $\chi\approx0.2$. Note that this range of shape factor values are generally where we see the sharpest gradients in the $S$-profiles plotted in figure~\ref{fig:TEXFRONTPROP:PROF} and also indicated by the superimposed contours in figure~\ref{fig:FULL:CF}. 

The minimum error is found at $S_{th}=0.85$ where the slope is $\kappa=1.89$. From the velocity field, we measure a front propagation factor in the transverse direction as $\kappa_{PIV}=1.96\pm0.4$. It should be noted that the speed estimated from the PIV is from the $0.5U_c$ contour, which is itself a surrogate, albeit a commonly used one \citep{Waitukaitis2012, Peters2014, Han2016, Peters2016, Majumdar2017, Han2018, Han2019, Baumgarten2019, Romcke2021a}.  Thus, our method shows that the propagation of the texture transition, and the propagation of the jamming front are comparable.


\subsection{Comparing texture and PIV data for the full field}\label{sec:Result:TEXCORR} 

Similarities between the texture measurements and the PIV data have been noted in the previous sections. Here, we address the full field of view explicitly. Most notably, the evolution of the region exhibiting surface features (that is $S<0.85$ in figure~\ref{fig:FULL:CF}) and the region traversing with the cylinder (that is $u/U_c>0.5$ in figure \ref{fig:DATA}e-h). In addition, it is observed that the surface features (figure \ref{fig:FULL:Q}) and the eigenvectors of the strain tensor ($\mathbf{n}_1$ and $\mathbf{n}_2$ in figure \ref{fig:DATA}i-l) are approximately oriented in the same directions. This section aims to quantify these observations. 

\begin{figure*}
    \includegraphics[width=\textwidth]{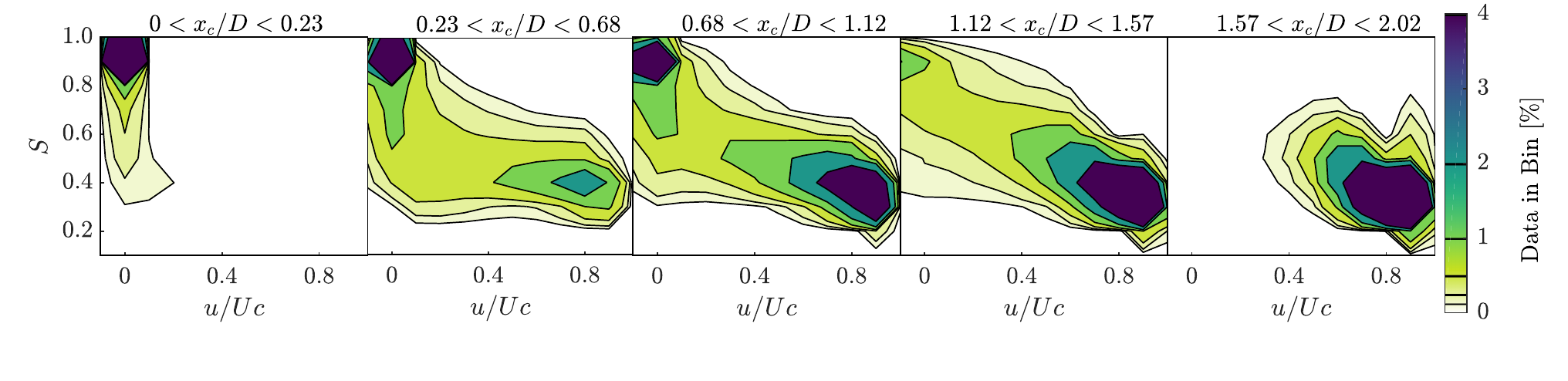}
   \caption{Evolution of the velocity, $u$, from the PIV experiment and shape factor, $S$, from the texture experiment. The figures are 2D histograms indicating where the data points tend to cluster for different stages in the experiment.}
   \label{fig:CFU} 
\end{figure*}

The combined evolution of the shape factor and velocity are plotted in figure \ref{fig:CFU}. As indicated in figure~\ref{fig:DATA} and \ref{fig:FULL:CF}, the overall trend of the data is to transition from a state of low velocity ($u/U_c<0.5$) and high shape factor ($S>0.85$) to a state of high velocity ($u/U_c>0.5$) and low shape factor ($S<0.85$). In other words, the suspension transitions from a quiescent suspension with no notable surface features, to moving with the cylinder while exhibiting observable surface features. Importantly, figure~\ref{fig:CFU} shows that the transition between these two states occurs at the same time in the experiment. 

In addition, we seek to rigorously confirm that the surface features and eigenvectors are oriented in the same direction. This analysis is only relevant where the suspension has deformed sufficiently and the surface exhibit clear surface features. As such, the data will be separated into regions where the analysis is conducted separately. 

\begin{figure}
  \centering
  \subcaptionbox{\label{fig:TEXCORR:CFUHIST}}{\includegraphics[width=0.7\columnwidth]{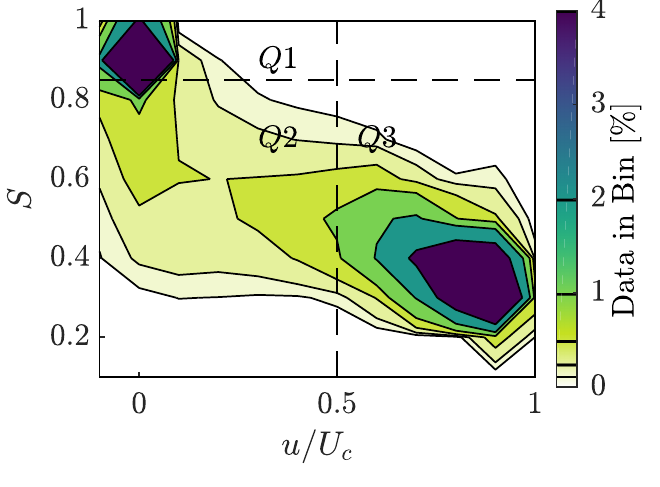}}\\
  \subcaptionbox{\label{fig:TEXCORR:QFIELD}}{\includegraphics[width=0.7\columnwidth]{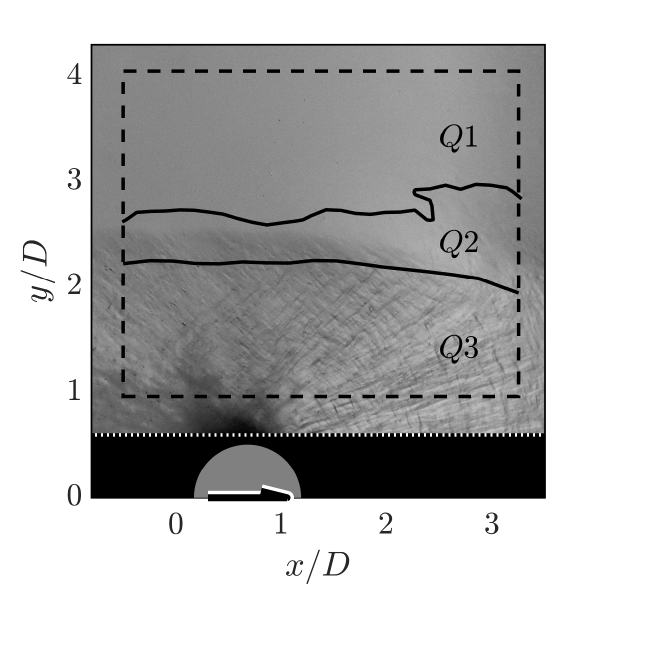}}
 \caption{(a) Histogram of $S$ and $u/U_c$ similar to figure \ref{fig:CFU} for the whole time series for all experimental runs. The dashed lines separates the data into quadrants $Q1$, $Q2$ and $Q3$. (b) Example of how the quadrants represent different regions of the system. This particular snapshot compares the velocity field and texture from figure~\ref{fig:DATA}f and \ref{fig:DATA}n.}
  \label{fig:TEXCORR}  
\end{figure}

As a first step towards identifying the relevant region, figure \ref{fig:TEXCORR:CFUHIST} shows the histogram of $S$ and $u/U_c$ similar to figure \ref{fig:CFU} for the full time series and all experimental runs. By using the threshold value $S_{th}=0.85$ for the shape factor (see section \ref{sec:Result:TEXPROP}) and the definition of the jamming front ($u/U_c=0.5$), we divide the data into quadrants. Here, $Q1$ represents a slow moving texture free state, $Q2$ represents the transition, while $Q3$ represents the suspension moving with the cylinder exhibiting measurable surface features. It is worth noting that $Q1$, $Q2$ and $Q3$ contain roughly $35$, $15$ and $50$~\% of the data, respectively. The unlabeled quadrant in figure \ref{fig:TEXCORR:CFUHIST} contains less than $1$~\% of the data, and its contribution is negligible. An example of where these regions are located in the flow is presented in figure \ref{fig:TEXCORR:QFIELD}. This figure compares the velocity field in figure \ref{fig:DATA}f with the result of the shape factor $S$ acquired from figure \ref{fig:DATA}n. 

\begin{figure}
  \centering
  \subcaptionbox{\label{fig:DOT:Schem}}{\includegraphics[width=0.6\columnwidth]{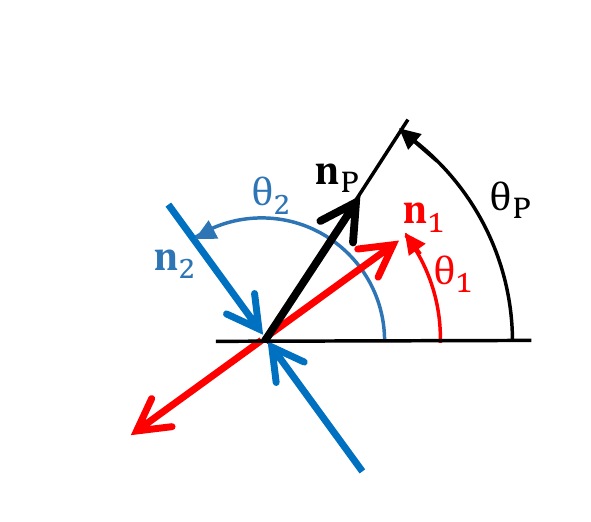}}\\\vspace{0.5em}
  \subcaptionbox{\label{fig:DOT:Corr}}{\includegraphics[width=0.9\columnwidth]{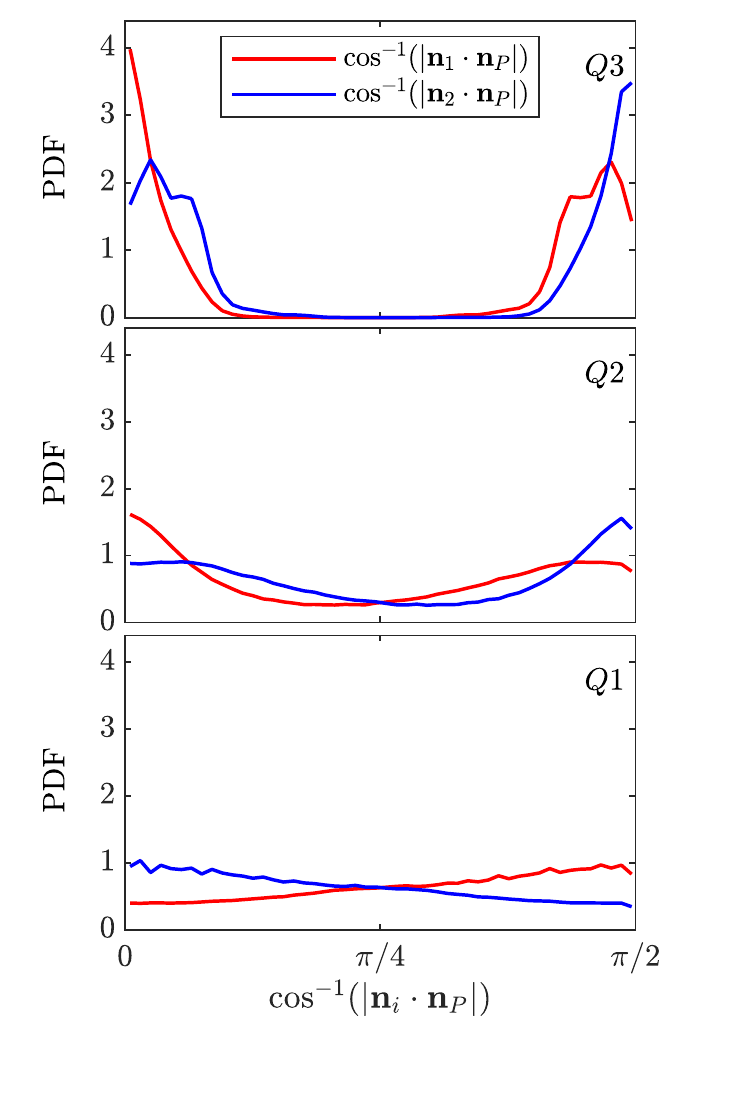}}
  \caption{(a) Schematic representation of texture orientation, $\mathbf{n}_P$, and strain eigenvectors where $\mathbf{n}_1$ represents stretch and $\mathbf{n}_2$ represents compression. (b) PDFs of the angle separating $\mathbf{n}_P$ and $\mathbf{n}_i$ for the regions $Q1$, $Q2$ and $Q3$, respectively.}
  \label{fig:AppDot} 
\end{figure}

A representation of the texture orientation and the strain eigenvectors are presented in figure \ref{fig:DOT:Schem}. The orientations are arbitrary, and the figure is only meant to be illustrative. The vectors are normalized to unit vectors. $\mathbf{n}_P$ is defined in section \ref{sec:Method:FFTSecAvg} and  represents the orientation of the texture. As a basis for comparing texture and eigenvector orientation we will use the dot products $\mathbf{n}_1\cdot\mathbf{n}_P$ and $\mathbf{n}_2\cdot\mathbf{n}_P$. Since all three vectors are unit vectors, these dot products represents the cosine of the angle separating them.

Figure \ref{fig:DOT:Corr} shows probability density functions (PDF) of $\cos^{-1}(|\mathbf{n}_i\cdot\mathbf{n}_P|)$ for the regions $Q1$, $Q2$ and $Q3$, respectively. The absolute value of the dot product is used here, thus representing the angle separating $\mathbf{n}_P$ and the span of the eigenvectors $\mathbf{n}_i$ with a positive value. As a result, $\cos^{-1}(|\mathbf{n}_i\cdot\mathbf{n}_P|)=0$ indicates $\mathbf{n}_P\parallel\mathbf{n}_i$, while $\cos^{-1}(|\mathbf{n}_i\cdot\mathbf{n}_P|)=\pi/2$ indicates $\mathbf{n}_P\perp\mathbf{n}_i$. As seen in figure \ref{fig:DOT:Corr}, the texture vectors are somewhat biased towards $\mathbf{n}_2$ in the $Q1$ region. In the transition region, $Q2$, the texture $\mathbf{n}_P$ starts to favor the $\parallel$ and $\perp$ directions relative to the strain. For $Q3$, the PDFs are $\approx0$ in the region $\pi/8<\cos^{-1}(|\mathbf{n}_i\cdot\mathbf{n}_P|)<3\pi/8$ with strong peaks at $0$ and $\pi/2$. In other words, the PDFs from region $Q3$ clearly show that the texture observed at the free surface has a strong connection with the orientation of the eigenvectors of the strain tensor.


\section{\label{sec:Conclusion} Conclusion}
Measurements of how surface texture evolves on a dense suspension of cornstarch and water with a freely propagating shear jamming front have been presented. The surface texture is captured by high-speed images of the free surface looking into a direct reflection. PIV measurements are used as a reference. From images of the surface texture, a 2D fast Fourier transform of local interrogation windows is used as a basis for analysing the surface structure. By taking a sector average of the power spectra, we are able to identify whether surface features are observed and the direction they are oriented. 

The PIV and texture measurements are two separate experiments, however, we show that the region of the suspension that shows clear surface features overlaps with the jammed region. In addition, we show that in the jammed, textured region of the flow, the eigenvectors of the strain tensor, and the observed surface features are oriented in the same direction.  Hence, our analysis reveals that pictures of the free surface contain quantifiable information not previously directly accessed.

Dilation \citep{Brown2012, Jerome2016, Majumdar2017, Maharjan2021} as well as surface corrugations \citep{Loimer2002, Timberlake2005} have been observed at the free surface of dense suspensions before. However, few studies have investigated the surface texture at a freely propagating shear jamming front \citep{Allen2018}.  The results presented in the current work, particularly the relation between the eigenvectors of the strain tensor and the orientation of the surface features, will provide insight for future model development and understanding of dense suspensions as well as a valuable measurement tool for future investigations.

\section*{Acknowledgements}
RJH is funded by the Research Council of Norway through project no.~288046. IRP acknowledges financial support from the Royal Society (Grant No.~RG160089).


\bibliographystyle{spbasic}      

\bibliography{REF}
\end{document}